\documentclass[iop]{emulateapj}
\usepackage{graphicx}
\shorttitle{Mechanical properties of dust aggregates probed by a solid-projectile impact}
\shortauthors{Katsuragi \& Blum}
\begin{document}

\title{The Physics of Protoplanetesimal Dust Agglomerates. X. Mechanical properties of dust aggregates probed by a solid-projectile impact}

\author{Hiroaki Katsuragi}
\affiliation{Institut f\"ur Geophysik und extraterrestrische Physik, Technische Universit\"at zu Braunschweig, Mendelssohnstr. 3, D-38106 Braunschweig, Germany}
\affiliation{Department of Earth and Environmental Sciences, Nagoya University, Furocho, Chikusa, Nagoya, Aichi 464-8601, Japan}

\author{J\"urgen Blum}
\affiliation{Institut f\"ur Geophysik und extraterrestrische Physik, Technische Universit\"at zu Braunschweig, Mendelssohnstr. 3, D-38106 Braunschweig, Germany}

\begin{abstract}
Dynamic characterization of mechanical properties of dust aggregates has been one of the most important problems to quantitatively discuss the dust growth in protoplanetary disks. We experimentally investigate the dynamic properties of dust aggregates by low-speed ($\lesssim 3.2$~m~s$^{-1}$) impacts of solid projectiles. Spherical impactors made of glass, steel, or lead are dropped onto a dust aggregate of packing fraction $\phi=0.35$ under vacuum conditions. The impact results in cratering or fragmentation of the dust aggregate, depending on the impact energy. The crater shape can be approximated by a spherical segment and no ejecta are observed. To understand the underlying physics of impacts into dust aggregates, the motion of the solid projectile is acquired by a high-speed camera. Using the obtained position data of the impactor, we analyze the drag-force law and dynamic pressure induced by the impact. We find that there are two characteristic strengths. One is defined by the ratio between impact energy and crater volume and is~$\simeq 120$ kPa. The other strength indicates the fragmentation threshold of dynamic pressure and is~$\simeq 10$ kPa. The former characterizes the apparent plastic deformation and is consistent with the drag force responsible for impactor deceleration. The latter corresponds to the dynamic tensile strength to create cracks. Using these results, a simple model for the compaction and fragmentation threshold of dust aggregates is proposed. In addition, the comparison of drag-force laws for dust aggregates and loose granular matter reveals the similarities and differences between the two materials.
\end{abstract}

\keywords{methods: laboratory ---
planetary system: formation --- planets and satellites: formation}



\section{Introduction} \label{sec:intro}
In protoplanetary disks, dust growth from sub-micrometer-sized monomer grains to at least kilometer-sized planetesimals has to occur in order to initiate planet formation. Once planetesimals have formed, their own gravity enables them to grow  towards planets in mutual collisions. However, the scenario of planetesimal formation is not so straightforward. According to \citet{Guttler:2010} and \citet{Zsom:2010}, the growth of dust aggregates by mutual adhesive collisions is limited to centimeter size, due to bouncing rather than the sticking collisions. Although their model was the first that compiled the results of numerous laboratory experiments~\citep[see][for an earlier compilation]{Blum:2008}, it was far from completion. For example, the influence of the size ratio of colliding aggregates~\citep{Syed:2017} or erosion~\citep{Schraepler:2017} have been experimentally studied only recently and revealed new aspects of dust-aggregate collisions.

In addition to the direct collision experiments, mechanical characterizations of bulk dust aggregates have also been performed. Compressive and tensile strengths of dust aggregates have been measured by static laboratory experiments~\citep{Blum:2004,Blum:2006}. The projectile-penetration test as well as compressive tests has also been performed~\citep{Guttler:2009}. Tensile strengths of dust aggregates were measured also by the Brazilian-disk test~\citep{Meisner:2012}. In \citet{Meisner:2012}, the packing-fraction dependence of the tensile strength and the elastic modulus based on sound-speed measurements were obtained as well. These mechanical properties are requisites for continuum-based numerical modeling of dust-aggregate collisions~\citep{Sirono:2004,Guttler:2009}. Furthermore, these mechanical properties provide guidelines to our understanding what happens in various complex phenomena caused by collisions of dust aggregates.

Typical values of tensile strengths obtained by the above-mentioned experiments are on the order of $1$~kPa. This strength value is actually close to that for cometary meteoroids~\citep{TrigoRodriguez:2006}. Besides, the typical packing-fraction value estimated from the bulk density of cometary nuclei is approximately $0.4$~\citep{Blum:2006}. This value is also in the range of typical packing-fraction values of dust aggregates used in laboratory experiments. Therefore, model studies with dust aggregates could be useful to discuss physical phenomena occurring on comets. To model cometary processes by using dust aggregates, proper understanding of their mechanical properties is necessary. Although wet granular matter, which can be a model for cohesive grains, shows similar tensile strengths on the order of $1$ kPa, its constituent-grain size is usually macroscopic~\citep{Scheel:2008,Herminghaus:2013}, with typical values around millimeter scale. Therefore, as long as we consider comets as more or less pristine objects~\citep{Fulle:2017}, dust aggregates can be regarded as suitable analogues of pristine materials to study cometary processes. Namely, the mechanical characterization of dust aggregates provides helpful information for both, planetesimal-formation and cometary-surface processes.

Since the dust aggregates used in experiments usually consist of micrometer-sized solid (SiO$_2$) monomers, they can be regarded as a certain class of granular matter. Granular matter is defined as a collection of dissipative solid grains. The main difference between dust aggregates and granular matter is the role of gravity. In macroscopic granular systems, gravity (body force) plays an essential role. However, monomers in dust aggregates are too small and too cohesive to be affected by gravity. Due to their mutual van der Waals attraction, dust aggregates can sustain their bulk shape even for small packing fractions. Contrastively, it is difficult to produce aggregates with small packing fractions using macroscopic granular matter. Under small packing-fraction conditions, macroscopic granular matter exhibits the gaseous rather than the solid state, even under microgravity conditions. Thus, dust aggregates are intrinsically different from conventional granular matter. What is the difference between granular matter and dust aggregates in terms of mechanical properties? To answer this question, impact tests might be useful.

Impact drag-force modeling has been used to characterize mechanical properties of granular matter~\citep{Katsuragi:2016}. For instance, the equation of motion of a solid projectile impacting onto a granular matter,
\begin{equation}
m\frac{d^2z}{dt^2} = mg - kz - m\frac{v^2}{d_1},
\label{eq:granular-drag}
\end{equation}
has been established based on experimental results \citep{Katsuragi:2007}. Here, $m$, $t$, $z$, and $v$ are the mass of the projectile, the elapsed time from the impact moment, the instantaneous penetration depth, and the instantaneous velocity of the projectile, respectively; $g$ represents the gravitational acceleration. The relation between the two parameters $k$, $d_1$ and the system properties (e.g., projectile density, projectile size, and granular friction) has also been experimentally revealed by~\citet{Katsuragi:2013}. For granular impacts, the depth-dependent term $kz$ in Eq.~(\ref{eq:granular-drag}) comes from the hydrostatic pressure of the granular target. For dust-aggregate impacts, the same form can be applied as long as the penetration depth is shallow (see Sect.~\ref{sec:drag-force}). The term $mv^2/d_1$ relates to the momentum transfer in both cases. Note that the vertically downward direction corresponds to the positive direction of $z$ in Eq.~(\ref{eq:granular-drag}). Actually, the granular impact dynamics depends on various conditions, such as gravity~\citep{Nakamura:2013,Altshuler:2014} and packing fraction of the granular matter~\citep{Umbanhowar:2010,Royer:2011}. Equation~(\ref{eq:granular-drag}) has to be improved to cover all these broader conditions. However, the accessible range of packing fractions for macroscopic granular matter is very limited as mentioned above. The impact drag force for very small packing fractions has not yet been studied well. By measuring the impact drag force of dust aggregates with small packing fraction and comparing that result with the granular impact case, we can further characterize the mechanical properties of both dust aggregates and granular matter.

In this study, we are going to examine the mechanical properties of dust aggregates by impact tests with solid projectiles. Specifically, the maximum penetration depth and the dynamic pressure during each impact event are computed and analyzed to characterize the dynamic strength of the dust aggregates. In particular, we discuss two kinds of strengths that characterize the mechanical properties of dust aggregates in the dynamic regime. Using our experimental results, a simple model to compute the compaction induced by the impact and the fragmentation threshold will be derived. At the same time, the drag force model of Eq.~(\ref{eq:granular-drag}) is used to explain the deceleration dynamics of penetrating projectiles. Then, similarities and differences between macroscopic granular matter and dust aggregates are discussed on the basis of our experimental data.

\section{Experiment} \label{sec:experiment}
\begin{figure}[ht!]
\begin{center}
%\plotone{fig1.eps}
\resizebox{0.45\textwidth}{!}{\includegraphics{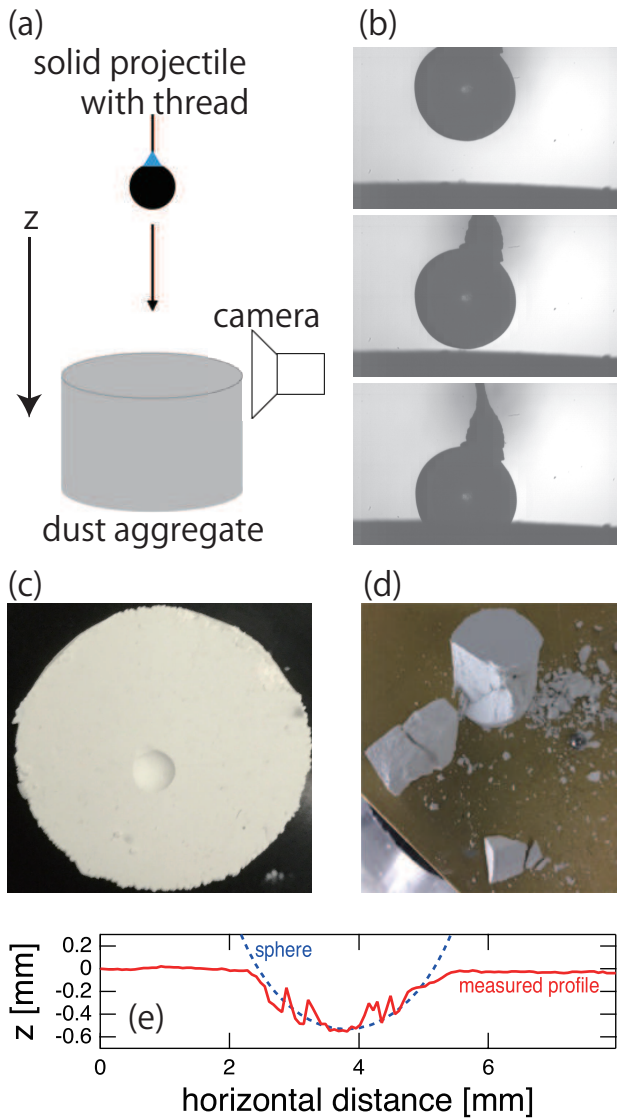}}
%\scalebox{1.0}[1.0]{\includegraphics{fig1.eps}}
\caption{Experimental setup and raw data. (a)~A solid projectile hung by a thread is released from a pre-deterimed height onto a dust aggregate in a vacuum chamber. (b)~ Side view images of the impact of a glass sphere of diameter $D=4.0$~mm with $v_0=2.0$~m s$^{-1}$. (c)~Top view of the crater created by the glass-sphere impact shown in (b). (d)~Fragmentation caused by an impact of a steel sphere of $D=4.0$~mm with $v_0=2.6$~m~s$^{-1}$. (e)~Comparison between the measured crater cross section (solid curve) and the spherical shape of the projectile (dashed curve). The crater is produced by an impact of a steel sphere with $D=4.0$~mm diameter and an impact velocity of $v_0=1.7$~m~s$^{-1}$. Mind the different scales of the two axes.}
\label{fig:setup}
\end{center}
\end{figure}

Our experimental setup is very simple and is shown in~Fig.~\ref{fig:setup}(a). A cylindrical dust aggregate of packing fraction $\phi=0.35$, diameter $20$~mm, and height $20$~mm is prepared as a target \citep[see][for a detailed description of the preparation method]{Blum:2014}. The cylindrical sample is made by using a mold and piston. A fixed amount of monodisperse spherical SiO$_2$ grains of radius $0.76~\mu$m is poured into a mold of $20$~mm in diameter. Then, it is compressed by a piston to $20$~mm in height and $\phi=0.35$. We employ monodisperse spherical monomers since they are ideally suited for understanding the physical mechanism and future comparison to particle codes. Although the exact mechanical properties of dust aggregates depend on the shape and size distribution of the constituent grains, their qualitative behavior is similar between aggregates consisting of monodisperse spherical grains and polydisperse irregular grains. The dependencies of mechanical and morphological properties of dust aggregates on monomer morphologies are basically less than one order of magnitude~\citep{Blum:2006,Bertini:2009}. Further details on the material properties of monomers and dust aggregates can be found in \citet{Blum:2004,Blum:2006}.

%% The "ht!" tells LaTeX to put the figure "here" first, at the "top" next
%% and to override the normal way of calculating a float position

Then, the solid projectile is dropped onto the dust-aggregate target in a vacuum chamber in the laboratory, i.e., under the influence of gravity. The residual pressure in the vacuum chamber is kept at $0.10$~Pa. We employ three kinds of projectiles made of glass, steel, and lead. A thin thread is glued on the top of the projectile to be held by a release mechanism, which guarantees a reproducible free fall with zero initial velocity. Diameter $D$ and density $\rho_p$ (including thread-mass effect) of the projectiles are ($D$, $\rho_p$)$=$($4.0$~mm, $2.6\times 10^3$~kg~m$^{-3}$) for glass, ($4.0$~mm, $7.7\times 10^3$~kg~m$^{-3}$) for steel, and  ($4.5$~mm, $11\times 10^3$~kg~m$^{-3}$) for lead. The impact velocity ranges from $v_0=0.19$~m~s$^{-1}$ to $v_0=3.2$~m~s$^{-1}$ and is controlled by the free-fall height of the projectile. In this study, $21$ impact experiments ($7$ with glass, $8$ with steel, and $6$ with lead) were carried out with various impact velocities. The impact kinetic energy $E=mv_0^2/2$ ranges from $E=3.6{\times}10^{-6}$~J to $E=1.1{\times}10^{-3}$~J.

The impact of the projectile is recorded by a high-speed camera (Photron SA-5) using a macro lens. The images of size $512 \times 320$ pixels with spatial resolution of $20$~$\mu$m~pixel$^{-1}$ are acquired with a rate of $42,000$ frames per second. Trimmed example images of the impacting projectile taken by the high-speed camera are shown in Fig.~\ref{fig:setup}(b). The position of the projectile is determined by image analysis. The image correlation of the upper hemisphere of the projectile is used to precisely identify the position of the projectile. Using the kinematic data obtained by the image analysis, we derive the mechanical properties of the target dust aggregate and drag-force law experienced by the impinging solid projectile, as will be shown below.

\section{Results and analyses} \label{sec:results}
\subsection{Cratering and fragmentation} \label{sec:crater-fragment}
When the kinetic energy of the impactor is small, only a shallow crater is formed on the target surface. An example of a crater is shown in Fig.~\ref{fig:setup}(c). To evaluate the crater morphology, laser profilometry is used, like in \citet{deVet:2007}. However, the accuracy of our simple system is not sufficient to derive the detailed properties of the crater, in particular its depth. Thus, we adopt the laser-profilometry method only to qualitatively confirm that the crater shape can be approximated by a spherical segment with the projectile's diameter. The comparison between the measured crater shape and the spherical shape of the projectile is shown in Fig.~\ref{fig:setup}(e). The good agreement of two profiles means that the crater is just an imprint of the impact. Neither excavation of target material nor ejecta splashing by the impact was observed. The crater is formed solely by the compression of the target dust aggregate. This type of cratering has not been found in other low-speed granular impact studies~\citep[see][]{Walsh:2003,Katsuragi:2010,PachecoVazquez:2011}. When the kinetic energy of the impact exceeds a certain threshold value, fragmentation of the target dust aggregate occurs. An example of a fragmented dust aggregate is shown in Fig.~\ref{fig:setup}(d).

\subsection{Kinematics of projectile motion} \label{sec:kinematics}
\begin{figure}[ht!]
\begin{center}
%\plotone{raw.eps}
\resizebox{0.45\textwidth}{!}{\includegraphics{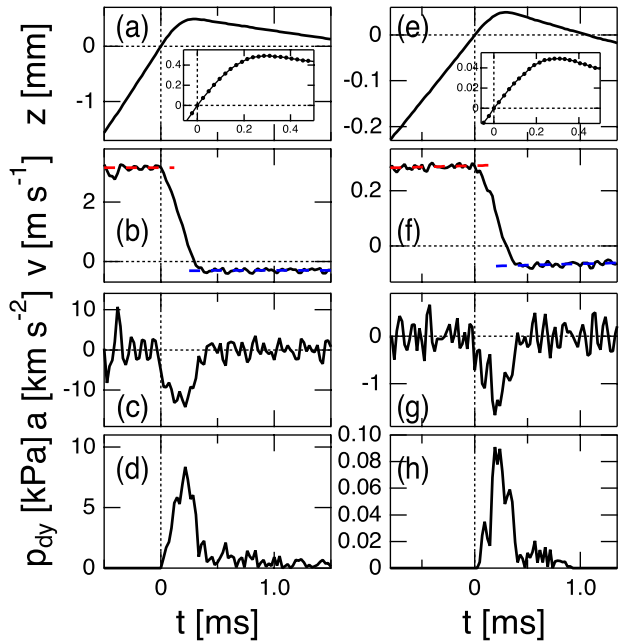}}
%\scalebox{1.0}[1.0]{\includegraphics{raw.eps}}
\caption{Example data of projectile motion during impact. The left and right column corresponds to glass-sphere impacts with impact velocities of $v_0=3.2$~m~s$^{-1}$ and $v_0=0.29$~m~s$^{-1}$, respectively. From top to bottom, penetration depth~$z$, velocity~$v$, acceleration~$a$, and dynamic pressure~$p_{\rm dy}$ are shown as a function of time $t$. In the insets of (a,e), $z(t)$ data during the penetration are magnified to display the individual data point. Units in the insets are identical to those used in the main plots. As normalization, we chose $t=0$ and $z=0$ for the impact moment and target surface, respectively. Although the difference of impact velocity between the left and right case is more than one order of magnitude, the penetration dynamics looks similar. Mind the units of km~s$^{-2}$ in the acceleration data.}
\label{fig:raw}
\end{center}
\end{figure}

Example data of the projectile motion during impact obtained by image analysis of the raw video data are shown in Fig.~\ref{fig:raw}. The left column exhibits a high-speed impact ($v_0=3.2$~m~s$^{-1}$), whereas the right column displays a low-speed case ($v_0=0.29$~m~s$^{-1}$). A glass-sphere projectile of $D=4.0$~mm is used in both cases. In the top row~(Fig.~\ref{fig:raw}(a,e)), the penetration depth $z$, which is directly measured from the image data, is shown as a function of time $t$. $z=0$ and $t=0$ correspond to the initial surface level of the dust aggregate and the impact moment, respectively. In the insets of Fig.~\ref{fig:raw}(a,e), the penetrating regimes are magnified so that individual data points can be discerned.

The impact moment is identified by using the velocity data $v(t)$ shown in the second row, Fig.~\ref{fig:raw}(b,f). The velocity data are computed by differentiating $z(t)$. Then, free-fall lines are fitted to the obviously free-falling regime. The free-fall portions of the trajectory with slopes $g=9.8$~m~s$^{-2}$ are shown as red dashed lines in Fig.~\ref{fig:raw}(b,f). The moment at which the velocity curve deviates from the free-fall line is defined as the impact moment ($t=0$). As shown in Fig.~\ref{fig:raw}(b,f), the velocity data exhibit a very sudden decrease after the impact. Thus, it is not difficult to identify the impact moment. The measurement uncertainty in this experiment is estimated by the uncertainty of each free-fall fitting. In the later analyses, this fitting uncertainty is used to calculate the corresponding errors with the error propagation method.

In many impact events, a rebound of the projectile is observed. The rebound moment and its velocity can also be identified by a free-fall fitting. The rebound velocity is defined by the velocity at which the projectile motion follows the free-fall line again. The rebounding free-fall lines are shown as blue dashed lines in Fig.~\ref{fig:raw}(b,f).

By differentiating $v(t)$ once more, the acceleration $a(t)$ can be obtained~(Fig.~\ref{fig:raw}(c,g)). The acceleration data are noisy, because the second derivative of the $z(t)$ data was used without any smoothing. Note that the units of the vertical axes in Fig.~\ref{fig:raw}(c,g) are km~s$^{-2}$. Thus, the fluctuation of $a(t)$ is much larger than $g$.
However, we can still confirm a clear peak in the deceleration curves.

Although the impact velocities vary by one order of magnitude between the left and right columns in Fig.~\ref{fig:raw}, the qualitative behavior of both impacts is quite similar. In both cases, a shallow cratering and rebound can be observed. Using steel and lead projectiles, fragmentation of dust aggregate was observed in $4$ impacts, while the remaining $17$ impacts resulted in the shallow cratering. In the fragmentation cases, the breaking target and deep penetration of the projectile prevented a reliable measurement by image analysis. Therefore, we mainly focus on the cratering regime in the following analyses.

\subsection{Penetration depth scaling} \label{sec:pnetration-depth}
\begin{figure}[ht!]
\begin{center}
%\plotone{d_H_scale.eps}
\resizebox{0.45\textwidth}{!}{\includegraphics{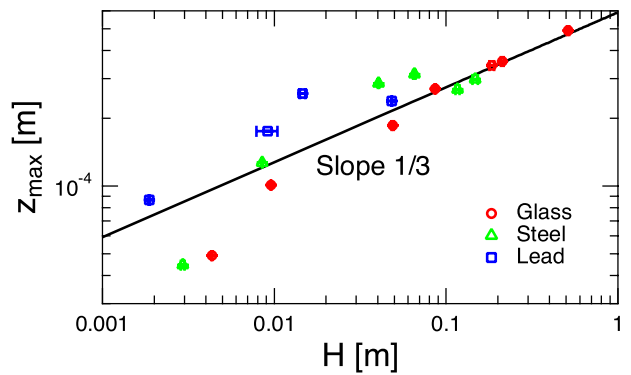}}
%\scalebox{1.0}[1.0]{\includegraphics{d_H_scale.eps}}
\caption{Scaling relation between the maximum penetration depth $z_{\rm max}$ and the total drop distance $H=h+z_{\rm max}$, where $h$ is the free-fall height. The relation $z_{\rm max}{\propto}H^{1/3}$ (straight line) can roughly be confirmed. Error bars are computed by error propagation from the uncertainties of the free-fall fitting.}
\label{fig:dH}
\end{center}
\end{figure}

From the peak value of $z(t)$, we can extract the maximum penetration depth. The maximum penetration depth corresponds to the crater depth since the crater is merely an imprint of the penetrating projectile as mentioned in Sect.~\ref{sec:crater-fragment} (see Fig.~\ref{fig:setup}(e)). The relation between the maximum penetration depth $z_{\rm max}$ and the total drop distance of the projectile $H=h+z_{\rm max}$, where $h$ is the free-fall height computed from the impact velocity through $h=v_0^2/2g$, is plotted in log-log style in Fig.~\ref{fig:dH}. The straight line in Fig.~\ref{fig:dH} shows the power-law relation $z_{\rm max}{\propto}H^{1/3}$. While the data show considerable scattering, they roughly agree with the power-one-third scaling. A similar power-law relation was found in experiments in which solid projectiles impacted into loose granular matter~\citep{Uehara:2003,Ambroso:2005}. Actually, this power-law relation simply implies that the penetration is in the strength regime~\citep{Katsuragi:2016}, which is atypical for coarse granular matter, but typical for cohesion-dominated material.

The scaling relation shown in Fig.~\ref{fig:dH} also implies that the impact energy governs the impact dynamics, since $H$ relates to the released potential energy. This tendency is different from the previous study by~\citet{Guttler:2009}, in which the impact between a glass sphere and a very porous dust aggregate ($\phi=0.15$) was experimentally investigated. \citet{Guttler:2009} found that the penetration depth can be scaled rather by momentum than by energy. In general, the momentum-dominant scaling explains the dissipative situation better. However, the current experimental result can be explained by energy scaling. In \citet{Guttler:2009}, deep penetration of the projectile was observed. On the contrary,  shallow penetration followed by rebound is the dominant outcome in our experiment. While both experiments are dissipative, the degree of dissipation might be different, which could result in the different scaling relations.

\subsection{Dynamic pressure} \label{sec:dynamic-pressure}
Here, we compute another quantity to characterize the dynamic properties of dust aggregates. To derive the dynamic strength of the dust aggregates, we estimated the dynamic pressure $p_{\rm dy}$. For shallow penetrations, the effect of contact between projectile and target must be taken into account to compute the instantaneous dynamic pressure.

\begin{figure}[ht!]
\begin{center}
%\plotone{zav0.eps}
\resizebox{0.45\textwidth}{!}{\includegraphics{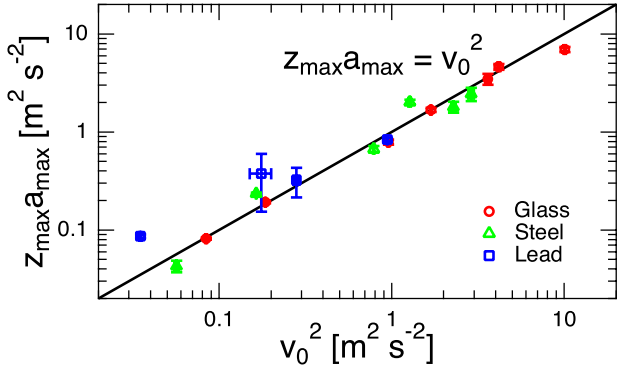}}
%\scalebox{1.0}[1.0]{\includegraphics{zab0.eps}}
\caption{Relation between $z_{\rm max}a_{\rm max}$ and $v_0^2$. The black line indicates $z_{\rm max}a_{\rm max}=v_0^2$ (Eq.~(\ref{eq:vza-rel})). Error bars are computed by the error propagation from the uncertainties of the free-fall fitting.}
\label{fig:za-max-v0}
\end{center}
\end{figure}

As mentioned in Sect.~\ref{sec:pnetration-depth}, the penetration dynamics is governed by the impact energy. Thus, let us begin with the energy conservation, i.e.,
\begin{equation}
\rho_p V_p a_{\rm max} \left( \frac{V}{A} \right)_{\rm max} = \frac{1}{2}\rho_p V_p v_0^2,
\label{eq:energy-conservation}
\end{equation}
where $V_p$ is the projectile volume and $a_{\rm max}$ is the maximum deceleration; $a_{\rm max}=\mbox{max}(|a|)$. $V$ and $A$ are the instantaneous penetration (crater) volume $V={\pi}(Dz^2/2-z^3/3)$ and the contacting cross-sectional area projected onto the initial target surface $A={\pi}z(D-2z)$, respectively. Thus, $(V/A)_{\rm max}$ denotes the maximum value of the effective penetration depth. By assuming a spherical crater shape in shallow penetration, $(V/A)_{\rm max}$ can be approximated as $z_{\rm max}/2$ by neglecting higher order terms. Thus, 
\begin{equation}
z_{\rm max}a_{\rm max} = v_0^2,
\label{eq:vza-rel}
\end{equation}
is obtained. This simple kinematic relation can be checked directly against the experimental data and is shown in Fig.~\ref{fig:za-max-v0}. As expected, all data agree very well with Eq.~(\ref{eq:vza-rel}), which is free from any fitting parameters.

To evaluate the dynamic pressure caused by the impact, we have to estimate the volume supporting the applied pressure. Actually, \citet{Guttler:2009} performed similar penetration experiments to ours and measured the compressed zone in the target dust aggregate by using the X-ray micro-CT method. According to their results, the volume of the compressed zone is equivalent to $0.8$~-~$1.2$ of the projectile-sphere volume. Therefore, we assume that the impact-affected volume corresponds to $V_p$. Then, the impact energy written in Eq.~(\ref{eq:energy-conservation}) is equivalent to ${\rm max}(p_{\rm dy})V_p$. From this assumption and Eqs.~(\ref{eq:energy-conservation}) and (\ref{eq:vza-rel}), we arrive at
\begin{equation}
\mbox{max}(p_{\rm dy}) =\frac{1}{2}\rho_p z_{\rm max}a_{\rm max}  =\frac{1}{2}\rho_p v_0^2.
\label{eq:dynamic-pressure2}
\end{equation}
Equation~(\ref{eq:dynamic-pressure2}) reveals the relations between the mechanical properties, kinematic values, and the initial conditions, by means of the dynamic pressure. To compute the instantaneous value of the dynamic pressure, $p_{\rm dy}$, we extend this relation to every instant by,
\begin{equation}
p_{\rm dy} = \frac{1}{2} \rho_p |a| z.
\label{eq:dynamic-pressure}
\end{equation}
This relation is geometrically equivalent to $p_{\rm dy}=\rho_p V |a|/A$ in shallow penetration. The computed values of $p_{\rm dy}$ are shown in Fig.~\ref{fig:raw}(d,h). Since both $a$ and $z$ vary by about one order of magnitude between the left and right columns in Fig.~\ref{fig:raw},  $p_{\rm dy}$ differs by about two orders of magnitude.

\subsection{Drag force characterization} \label{sec:drag-force}
\begin{figure*}[ht!]
\begin{center}
%\plotone{v_z_f.eps}
\resizebox{0.9\textwidth}{!}{\includegraphics{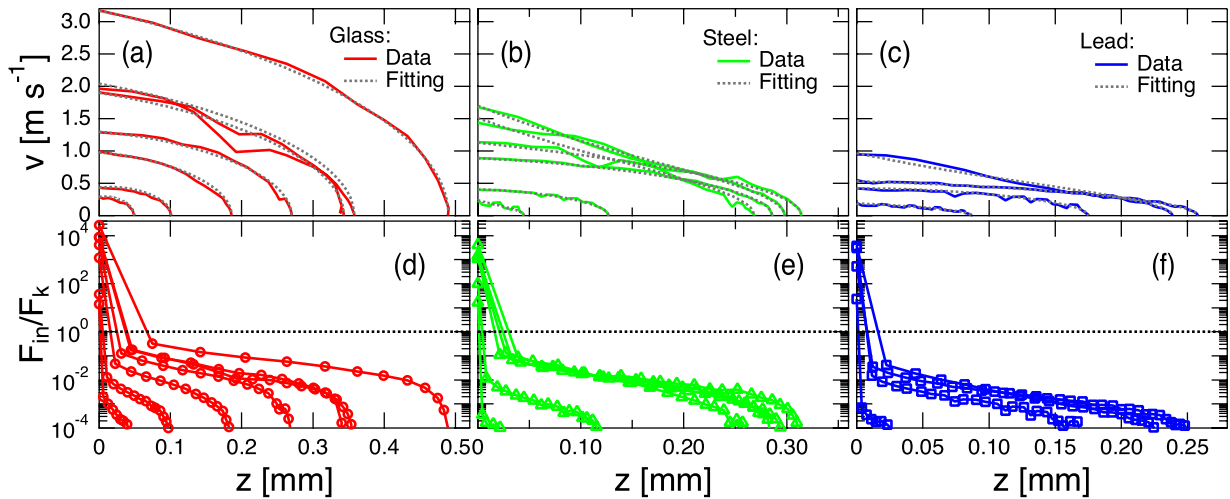}}
%\scalebox{1.0}[1.0]{\includegraphics{v_z_f.eps}}
\caption{Top row: instantaneous velocity $v$ as a function of the penetration depth $z$ of (a)~glass, (b)~steel, and (c)~lead projectiles with various impact energies. Colored curves indicate the experimental results while the gray dashed curves show fitting results. Bottom row: ratio between inertial drag $F_{\rm in}$ and deformation-based drag $F_k$ for each impact ((d)~glass, (e)~steel, and (f)~lead) computed from the fitting results. Except for the very early stage of impact, $F_{\rm in}$ is much smaller than $F_k$. }
\label{fig:vzf}
\end{center}
\end{figure*}

To analyze the penetration dynamics, we tried to fit the experimental data to the simple drag-force model written in Eq.~(\ref{eq:granular-drag}). Equation~(\ref{eq:granular-drag}) has an analytic solution in $v$-$z$ space~ \citep[Eq.~(2) in][]{Katsuragi:2013},
\begin{equation}
\frac{v^2}{v_0^2} = e^{-\frac{2z}{d_1}}-\frac{k d_1 z}{m v_0^2}
   +\left( \frac{g d_1}{v_0^2}+\frac{k d_1^2}{2 m v_0^2}\right) \left(1- e^{-\frac{2z}{d_1}} \right).
\label{eq:solution}
\end{equation}
Using this analytic solution, the experimentally obtained $v(z)$ data can be fitted. The fitting results are shown in Fig.~\ref{fig:vzf}(a-c). The colored curves are the measured data and the gray dashed curves are the fit results. Here, the two parameters $k$ and $d_1$ were taken as free fit parameters. As shown in Fig.~\ref{fig:vzf}(a-c), all experimental data can be very well fitted by Eq.~(\ref{eq:solution}). Note that the values of $k$ and $d_1$ do not depend on time $t$, because they are determined by fitting the entire shape of each curve shown in Fig.~\ref{fig:vzf}(a-c). The values of the two fit parameters $k$ and $d_1$ provide helpful information to discuss the physical situation of the impact.

\begin{figure}[ht!]
\begin{center}
%\plotone{k_d1.eps}
\resizebox{0.45\textwidth}{!}{\includegraphics{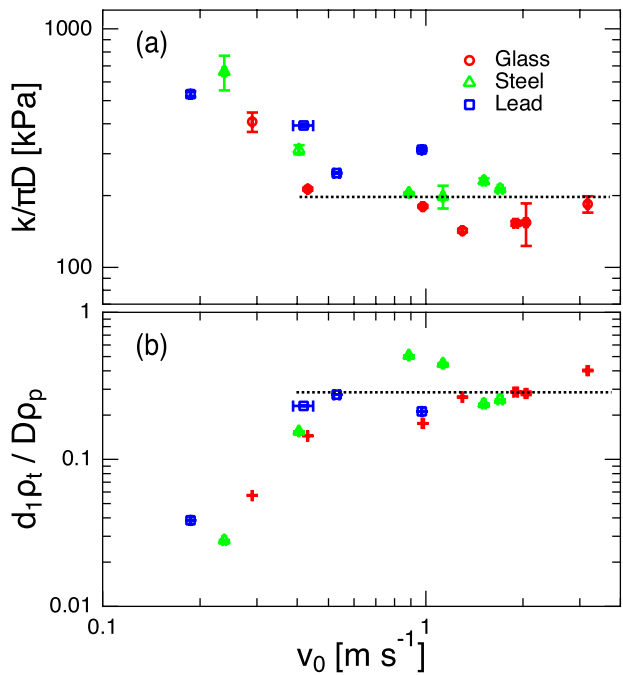}}
%\scalebox{1.0}[1.0]{\includegraphics{k_d1.eps}}
\caption{Normalized fit parameter values characterizing (a)~deformation-based drag $k/({\pi}D)$ and (b)~inertial drag $d_1~{\rho_t}/({\rho_p}~D)$. For relatively high impact velocities ($v_0 \gtrsim 0.4$ m s$^{-1}$), these quantities become practically constant. Horizontal dashed lines are intended to guide the eye. Error bars for $k/({\pi}D)$ and $d_1~{\rho_t}/({\rho_p}~D)$ indicate the uncertainty of fitting shown in Fig.~\ref{fig:vzf}(a-c). Error bars for $v_0$ are estimated from the uncertainties of the free-fall fitting.}
\label{fig:k-d1}
\end{center}
\end{figure}

Since the first drag-force term $F_k=kz$ is proportional to the penetration depth $z$, it relates to the deformation. For the impacts into granular matter, $F_k$ is related to the frictional force by the depth-proportional hydrostatic pressure and Coulomb friction of the granular target. However, the hydrostatic pressure is irrelevant for impacts into cohesive dust aggregates. Rather, elastic and plastic deformations should determine $F_k$ in this case. Assuming that the impact energy is dissipated by plastic deformation forming the cratering imprint, the deformation-based force $F_k=kz$ should be related to the strength. To compute the relevant strength, $F_k$ must be divided by the contact area. The contact area of a shallow  penetration of a sphere can approximately be written as $A{\simeq}{\pi} z D$ by neglecting the $z^2$ term. Then, the analogue of the strength can be estimated as $F_k/A{\simeq}k/({\pi}D)$. In Fig.~\ref{fig:k-d1}(a), $k/({\pi}D)$ is plotted as a function of the impact velocity $v_0$.

The second fit parameter $d_1$ characterizes the inertial-drag force $F_{\rm in}=m~v^2/d_{1}$, which originates from the momentum transfer between projectile and target. The momentum transfer can be written as $\rho_t ~ v^2 ~ D^2 ~ dt \sim \rho_p ~ D^3 ~dv$, where $\rho_t$ and $\rho_p$ are the bulk densities of target and projectile, and $\rho_p~D^3~dv/dt=F_{\rm in}$ is the inertial-drag force based on the infinitesimal deceleration $dv$ during the infinitesimal duration $dt$. From this momentum balance, a relation $d_1{\sim}(\rho_p/(\rho_t)~D)$ can be derived. To characterize the inertial drag by a dimensionless number corresponding to a {\em drag coefficient}, we plot $d_1{\rho_t}/({\rho_p}~D)$ in Fig.~\ref{fig:k-d1}(b). Here, the contact area is approximated by $D^2$ for the sake of simplicity. Although the actual contact area is proportional to ${\pi} z D$, the factor ${\pi} z$ is replaced by a constant characteristic length scale $D$. Since the $F_{\rm in}$-dominant regime is limited to very shallow indentations as discussed later~(Fig.~\ref{fig:vzf}(d-f)), $z$ can be approximated with a constant. Thus, we replace ${\pi} z$ with $D$ in this case. Although $D$ is much greater than $\pi z$, there is no other relevant constant length scale in the system. While the specific value of normalized $d_1$ depends on the choice of the length scale, its qualitative behavior is similar as long as the constant length scale is used.

In Fig.~\ref{fig:k-d1}, both quantities $k/({\pi}D)$ and $d_1~{\rho_t}/({\rho_p}~D)$ become practically constant for  relatively high impact velocites ($v_0 \gtrsim 0.4$ m s$^{-1}$). The horizontal dashed lines in Fig.~\ref{fig:k-d1} are displayed to guide the eye. The saturated constant values are $k/({\pi}D){\simeq}200$ kPa and $d_1~{\rho_t}/({\rho_p}~D){\simeq}0.3$. The former value is significantly larger than the typical value of the dynamic strength shown in Fig.~\ref{fig:raw}(d,h). The reason for this difference is an important point to characterize dust-aggregate impacts as discuss later in Sect.~\ref{sec:strength-values}. The latter value is close to unity. This means that we appropriately characterize the inertial drag by means of momentum transfer. Moreover, the value of $d_1~{\rho_t}/({\rho_p}~D)$ is similar to that for the granular impact case~\citep{Katsuragi:2007,Katsuragi:2013}. Values slightly smaller than unity originate in using the length scale $D$ in the normalization of $d_1$. However, $D$ is the only relevant length scale in the localized impact compression.

The ratio of inertial and deformation-based drag forces, $F_{\rm in}/F_k$, can also be computed from the fitting result. The computed ratio is shown in Fig.~\ref{fig:vzf}(d-f). Obviously, the ratio exceeds unity only in the very early stages of impact. In fact, only the first data point shows large $F_{\rm in}/F_k$ values in all cases. Therefore, we can safely say that the inertial drag is efficient only within the initial $24$~$\mu$s ($=1/42,000$~s; temporal resolution of the measurement). After the very initial inertial-dominant regime, the dynamics is mainly governed by $F_k$. Therefore, $F_k$ must be the relevant parameter to estimate the penetration depth and resultant crater shape. This tendency is consistent with the strength-dominant cratering (penetration) process if we consider that $F_k$ characterizes plastic deformation.

\subsection{Scaling of crater volume, restitution coefficient, and dynamic pressure} \label{sec:crater-volume}
\begin{figure}[ht!]
\begin{center}
%\plotone{V_e_Mp.eps}
\resizebox{0.45\textwidth}{!}{\includegraphics{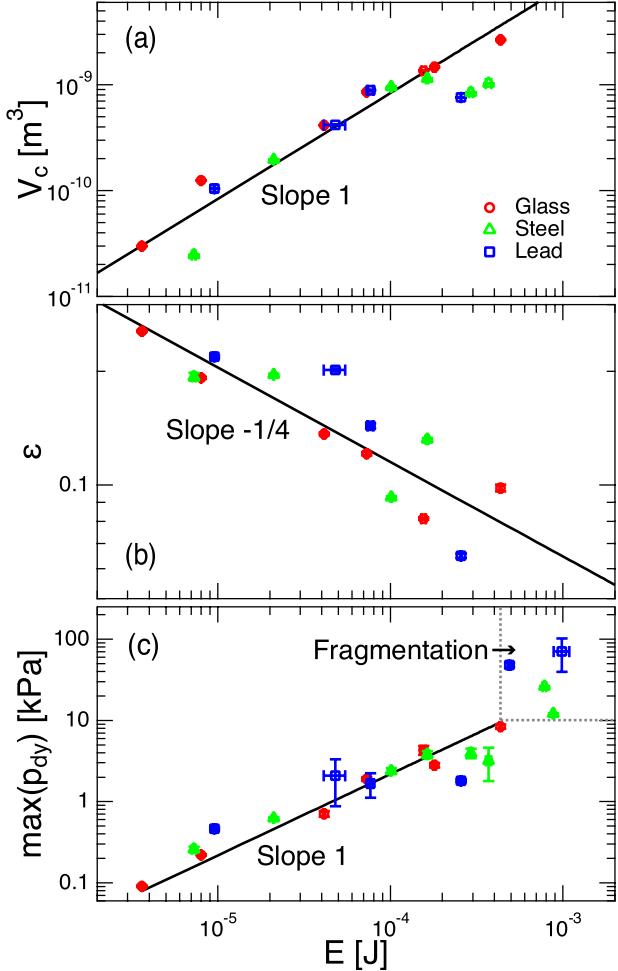}}
%\scalebox{1.0}[1.0]{\includegraphics{V_e_Mp.eps}}
\caption{Crater volume $V_c$ (a), restitution coefficient $\epsilon$ (b), and maximum dynamic pressure max($p_{\rm dyn}$) (c) as a function of kinetic impact energy $E$. (a)~The linear relation $V_c{\propto}E$ can be confirmed and the fitting quality is better than in Fig.~\ref{fig:dH}. (b)~The restitution coefficient $\epsilon$ can be scaled as $\epsilon \propto E^{-1/4}$. (c)~The relation max($p_{\rm dyn}$)${\propto}E$ is held in the cratering regime. Fragmentation of dust aggregates occurs at the highest impact energies (or equivalently at the largest dynamic pressures) beyond the dashed line levels. Error bars are computed by error propagation from the uncertainties of the free-fall fitting.}
\label{fig:Vep}
\end{center}
\end{figure}

In the strength regime, the crater volume should be proportional to the kinetic energy $E$ of the impact. Assuming a spherical crater shape, we can compute the crater volume $V_c={\pi}(Dz_{\rm max}^2/2-z_{\rm max}^3/3)$. The relation between $V_c$ and $E$ is plotted in Fig.~\ref{fig:Vep}(a) and shows a clear linear relation with the slope of unity in the log-log plot. Thus, the data suggest the relation
\begin{equation}
V_c =Y_{\rm k}^{-1} E,
\label{eq:vol-scale}
\end{equation}
where $Y_{\rm k}$ is an effective strength (yield strength) characterizing the cratering dynamics. We find that the quality of scaling by Eq.~(\ref{eq:vol-scale}) shown in Fig.~\ref{fig:Vep}(a) is better than the scaling by $z_{\rm max}{\propto}H^{1/3}$ shown in Fig.~\ref{fig:dH}. This improvement of scaling quality suggests that the assumption of a spherical crater is reasonable. From the least-square fitting, the strength value is obtained as $Y_{\rm k}= 120$~kPa. This value agrees well with the drag-based strength value, $k/({\pi}D)=200$~kPa~(in the saturated regime of Fig.~\ref{fig:k-d1}(a)).

If the collision is completely inelastic without any rebound, these strength values originate only from the pure plastic deformation and all the impact energy is dissipated by the cratering. However, we do observe rebound of the projectile in many impacts. For instance, rebounds can clearly be confirmed in Fig.~\ref{fig:raw}. As mentioned in Sect.~\ref{sec:kinematics}, the impact and rebound velocities $v_0$ and $v'$ can be determined by the data fitting. Then, the restitution coefficient $\epsilon=|v'/v_0|$ can be computed. The derived values of $\epsilon$ as a function of $E$ are shown in Fig.~\ref{fig:Vep}(b). In some cases, detection of rebound is impossible due to very small $v'$ or other limitations of image analysis. All impacts for which we could identify a rebound are plotted in Fig.~\ref{fig:Vep}(b). As can be seen, $\epsilon$ is clearly a decreasing function of $E$. The straight line in Fig.~\ref{fig:Vep}(b) indicates a power-law relation of
\begin{equation}
\epsilon \propto E^{-1/4}.
\label{eq:restitution}
\end{equation}
This relation actually conflicts with the classical theory for the restitution coefficient associated with plastic deformation~\citep{Johnson:1985}. While \citet{Johnson:1985} derived the relation $\epsilon \propto v_0^{-1/4}$, the current experimental result suggests $\epsilon \propto E^{-1/4} \propto v_0^{-1/2}$, considerably steeper as for classical elastic-plastic materials. The reason for this discrepancy is not clear at present. The complex rheological properties of dust aggregates may be responsible for this restitution behavior shown in Fig.~\ref{fig:Vep}(b) and described by Eq.~(\ref{eq:restitution}).

Finally, we analyze the dynamic pressure $p_{\rm dy}$. The maximum value of the dynamic pressure max($p_{\rm dy}$) could play a crucial role to characterize the mechanical properties of dust aggregates in a dynamic situation. In Fig.~\ref{fig:Vep}(c), we plot max($p_{\rm dy}$) as a function of $E$. Since the data suggest a linear dependence between the two values, we can write a relation
\begin{equation}
{\rm max(}p_{\rm dy}{\rm )} =V_s^{-1} E,
\label{eq:max-p}
\end{equation}
where $V_s$ is a constant having the dimension of a volume. From the least-squares fitting, we get $V_s=4.6{\times}10^{-8}$~m$^3$. This volume is much greater than the typical crater volume~${\simeq}10^{-9}$~m$^3$ and much smaller than the entire target volume~${\simeq}6.3{\times}10^{-6}$~m$^3$. Rather, $V_s$ is comparable to the typical projectile volume~$V_p{\simeq}3.3{\times}10^{-8}$~m$^3$. This relation $V_s \simeq V_p$ is natural since we have assumed that the volume supporting $p_{\rm dy}$ corresponds to $V_p$ (see Sect.~\ref{sec:dynamic-pressure}).

The threshold for fragmentation can also be derived by considering max($p_{\rm dy}$). The estimated max($p_{\rm dy}$) values for the fragmentation cases distribute in the top-right corner of Fig.~\ref{fig:Vep}(c) bounded by the dashed lines. This implies that, once max($p_{\rm dy}$) exceeds a certain threshold value, fragmentation occurs due to macroscopic crack propagation in the target dust aggregate. The threshold value for our dust-aggregate targets is $p_{\rm dy}^*{\simeq}10$~kPa~(Fig.~\ref{fig:Vep}(c)). This value is approximately one order of magnitude less than $k/({\pi}D)$ and $Y_{\rm k}$.

Note that this fragmentation threshold is not necessarily identical to the catastrophic disruption limit which is usually characterized by a largest fragment mass smaller than one half of the original mass. Although we do not measure the mass of the largest fragment, it is seemingly not always less than one half of the original mass so that $p_{\rm dy}^*$ denotes the onset of fragmentation.

\section{Discussion} \label{sec:discussion}
\subsection{Physical meaning of the strength values} \label{sec:strength-values}
The tensile strength of dust aggregates has been determined by considering the stress required to open a crack~\citep{Blum:2006}. By counting the number of monomer-monomer contacts per unit area in a dust aggregate, \citet{Blum:2006} obtained a form for the tensile strength $Y$ \citep[Eq.~(6) in][]{Blum:2006},
\begin{equation}
Y = \frac{3 \phi F_{\rm stick}}{2 \pi s_0^2},
\label{eq:tensile-strength}
\end{equation}
where $F_{\rm stick}$ and $s_0$ are the adhesion force and radius of monomers composing the dust aggregate, respectively. Inserting typical values of $\phi=0.35$, $F_{\rm stick}=67{\times}10^{-9}$~N~\citep{Heim:1999}, and $s_0=0.76$~$\mu$m~\citep{Poppe:2005} into Eq.~(\ref{eq:tensile-strength}), we obtain $Y=19$~kPa. This value is quite close to the fragmentation threshold derived by the dynamic pressure $p_{\rm dy}^* \simeq 10$~kPa. This correspondence is reasonable, because both quantities $Y$ and $p_{\rm dy}^*$ characterize the local criterion to open a crack. Once the macroscopic crack is opened in a dust aggregate, it will be unstable in a dynamic state and further propagate until the aggregate splits. Therefore, fragmentation takes place when the dynamic pressure exceeds the critical value $p_{\rm dy}^*$ in localized impacts. However, measured values of the {\it static} tensile strength for dust aggregates are less than the above-estimated value~\citep{Blum:2006,Meisner:2012}. To explain this difference, \citet{Blum:2006} have used the energy balance as well, which then yields a much smaller value for the tensile strength. However, the current experimental results suggest that the force balance is essential for estimating the tensile strength in the dynamic impact situation.

From the linear relation between max($p_{\rm dy}$) and the impact kinetic energy $E$, a volume $V_s (\simeq V_p)$ can be estimated as written in Eq.~(\ref{eq:max-p}), which corresponds to the volume that is supporting the applied stress max($p_{\rm dy}$). This agreement of $V_s$ and $V_p$ is physically reasonable from the viewpoint of dimensional analysis. If the size of the target is large compared to the projectile size, the relevant length scale, which determines the extent of the impact, can only be the size of projectile. Then, the volume of the impact-affected (compressed) zone naturally corresponds to the projectile volume. A similarly affected zone can actually be observed in a horizontally-dragged two-dimensional granular layer~\citep{Takehara:2010}.

In our experiments, the target volume is much greater than the projectile volume. As already mentioned, \citet{Guttler:2009} reported that the impact-induced compression is limited to a region approximately the volume of the projectile. That is, only the vicinity of the impact point is compressed. In addition, it has been shown that the boundary effect even in granular-impact-cratering experiment is very limited~\citep{Nelson:2008}. Thus, we consider the impact compression is localized to a volume much smaller than our target and the open-boundary effect is negligible at least in the cratering regime. However, the fragmentation limit certainly depends on the size of the target. Systematic experiments like those by \citet{Syed:2017} are necessary to reveal the size dependence of the fragmentation limit. In this study, we rather focus on the fragmentation limit using identical-size impactors in terms of the maximum dynamic pressure. Systematic studies with target-size variations are interesting future problems.

Recently, a periodic propagation of a localized compaction band has been found in a compressed crushable granular material~\citep{Valdes:2011,Guillard:2015}. The occurrence of a periodic compaction depends on the competition among elastic, local braking, and viscous timescales. Thus, if we can observe a similar oscillation with dust aggregates by controlling the compression rate, its analysis might provide further mechanical information. Numerical and experimental studies on the behavior of compressed zones are interesting future topics.

In this study, we also found another type of strength by the analyses of deformation, i.e., Eq.~(\ref{eq:vol-scale}) and the drag force fitting to Eq.~(\ref{eq:solution}). The two obtained values $Y_k=120$~kPa and $k/({\pi}D)=200$~kPa are close to each other and about one order of magnitude greater than the dynamic pressure threshold $p_{\rm dy}^*=10$~kPa. This is because the former two strengths are computed based only on the apparent deformation, whereas the latter takes into account the compressed zone beneath the crater as well. In other words, one must consider not only the direct deformation, like cratering, but also the compression of the surrounding zone to properly model the strength of very soft target materials, such as dust aggregates. However, the projectile motion and the resultant crater shapes are much easier to measure than the internal compression caused by the impact. To estimate the impact condition from the resultant crater shape, the effective strength $Y_k$ is convenient. As long as the scaling of Eq.~(\ref{eq:vol-scale}) is held, the cratering process can be understood effectively by the strength-dominant dynamics governed by the impact energy. Strength values obtained in this study are summarized in Table~\ref{tab:values}.

\begin{deluxetable}{ccl}[b!]
\tablenum{1}
\tablecaption{Strength values obtained or estimated in this study} \label{tab:values}
\tablecolumns{3}
\tablewidth{0pt}
\tablehead{
\colhead{Symbol} &
\colhead{Value (kPa)} &
\colhead{From}
}
\startdata
$Y_k$ & 120 & Crater shape, Eq.~(\ref{eq:vol-scale}) \\
$k/{\pi}D$ & 200 & Penetration dynamics, Fig.~\ref{fig:k-d1}(a)  \\
$p_{\rm dy}^*$ & 10 & Fragmentation threshold of $p_{\rm dy}$, Fig.~\ref{fig:Vep}(c)  \\
$Y$ & 19 & Stress to open a crack, Eq.~(\ref{eq:tensile-strength}) \\
\enddata
\end{deluxetable}

The relation between max($p_{\rm dy}$) and $Y_k$ can be derived from Eqs.~(\ref{eq:vol-scale}), (\ref{eq:max-p}) and $V_s{\simeq}V_p={\pi}D^3/6$, $Y_k$, and max($p_{\rm dy}$) in the following way
\begin{equation}
Y_k = \frac{1}{6}\left( \frac{D}{z_{\rm max}} \right)^2 \rho_p v_0^{2} = \frac{1}{3}\left( \frac{D}{z_{\rm max}} \right)^2 \mbox{max}(p_{\rm dy}).
\label{eq:Yk-pdy}
\end{equation}
Using this form and Eq.~(\ref{eq:dynamic-pressure2}), we can caluclate the strength value from the crater depth $z_{\rm max}$ and the impact velocity $v_0$. To apply these forms, the projectile diameter $D$ and its density $\rho_p$ must be constant, i.e., the projectile has to be much harder than the target. Then, the value of $Y_k$ can be estimated even from a single penetration test. To derive the fragmentation threshold $p_{\rm dy}^*$, however, one has to search for the threshold impact velocity $v_0^*$ at which the fragmentation begins to occur. Once we obtain $v_0^*$, the strength can be estimated by Eq.~(\ref{eq:dynamic-pressure2}).

To build a model unifying these strength values, systematic studies by experiments and numerical simulations are still needed. We have fixed the packing fraction of the dust aggregates in this experiment. In addition, our experiments were performed only under the influence of gravity. Variations of packing fraction and gravitational effect are crucial research topics to improve the model.

%%%%% HERE JB 20170907

\subsection{Possible applications in astrophysics} \label{sec:possible-effect}
In our experiments, we used dense solid projectiles, whereas in previous studies, mutual collisions of similar dust aggregates have mainly been investigated to directly simulate the planetesimal formation process. Collisions between objects of very different densities in protoplanetary disks might be not be the rule, they still might happen, e.g., when dust aggregates collide with chondrules \citep{Beitz:2013} or CAIs. Recently, \citet{Syed:2017} have experimentally shown that the size ratio of colliding dust aggregates is an important factor for the collisional outcome. In addition to the size ratio, the density ratio of two colliding bodies might also affect the result of mutual collisions. For example, in more energetic impact-cratering experiments with a dust-aggregate projectile, ejecta coming from all over the target surface were observed~\citep{Wurm:2005}. However, we did not recognize such ejection in our experiments. In the high impact-energy regime, fragmentation of the target is induced rather than the enhanced ejection of dust. To understand this difference, the influence of projectile density and structure have to be studied.

Since this work aims at discussing the fundamental mechanical properties of dust aggregates in the dynamical state, we employed a solid-projectile impact as a kind of the simplest test case. Although the collision outcomes could depend on the size and/or density ratio of the colliding bodies, we believe that the energy-based descriptions, like Fig.~\ref{fig:Vep}, and the correct strength values should basically be universal in terms of local dynamics around the impact point. Using our current result, we can estimate the deformation and volume of the compressed zone around the impact point. The specific model to calculate the compression and the onset of fragmentation will be discussed in the next section~(Sect.~\ref{sec:modeling}).

Repeated impacts of the type of impacts studied here may cause surface compaction and would render the target aggregate into a soft core with a hard shell around, as studied by \citet{Weidling:2009}. Such a process could induce a history dependence in dust-aggregate growth. Further systematic studies are necessary to discuss the importance or unimportance of surface compaction due to the low-energy impacts.

As for cometary, asteroidal or lunar surface processes, it is known that their surfaces experience various impacts with objects of different size, density and impact velocity, as well as the fall-back of excavated material that leads to the formation of loose regolith layers. To properly simulate the formation and compaction of regolith surfaces, a good knowledge of the impact physics into granular matter is required. Examples of applications of experimental impact studies such as the one described in this paper are high-velocity impacts into asteroidal surfaces with and without regolith \citep[see, e.g.,][]{Beitz:2016} or dust-aggregate impacts into granular matter \citep[see, e.g.,][]{Planes:2017}.

\subsection{Modeling of compaction and fragmentation of dust aggregates} \label{sec:modeling}
By using the results obtained so far, we propose a simple model to estimate the fragmentation threshold and surface compaction induced by an impact into a dust aggregate. First, the fragmentation-threshold strength of dust aggregates can be estimated by Eq.~(\ref{eq:tensile-strength}). If the maximum dynamic pressure computed by Eq.~(\ref{eq:dynamic-pressure2}) exceeds this fragmentation threshold, fragmentation occurs for homogeneous dust aggregates. From the current results, we cannot estimate the largest fragment mass and fragment-mass distribution. However, these properties were derived by \citet{Syed:2017} for aggregate-aggregate collisions. If the maximum dynamic pressure is less than the fragmentation threshold, compaction is induced. In these impact cases, a volume corresponding to the projectile volume $V_p$ is compressed to $V_p-V_c$, the compaction proceeds by a factor $V_p/(V_p-V_c)$. Equation~(\ref{eq:vol-scale}) provides the recipe to compute $V_c$. Although the value $Y_k=120$~kPa can be used for now, it is truly applicable for $\phi=0.35$ only. The dependence of $Y_k$ on the volume filling factor must be revealed in future studies to complete the compaction modeling. The simplest approximation is $Y_k(\phi){\simeq}120\phi/0.35$~(kPa). If the geometrical condition at the fragmentation threshold $D/z_{\rm max}^*$ in Eq.~(\ref{eq:Yk-pdy}) is independent of $\phi$, this approximation is plausible, because $p_{\rm}^*$ is proportional to $\phi$ in Eq.~(\ref{eq:tensile-strength}). Then, $V_c$, which is a key factor to calculate the compaction, can be estimated with Eq.~(\ref{eq:vol-scale}). By assuming many random impacts, patchy-compacted surfaces result in an inhomogeneity of the dust aggregates. To discuss the collision outcomes of inhomogeneous dust aggregates, more complex (history-dependent) impact tests are required both by experiments and numerical simulations. Our study provides only the first-step of modeling for such complex modes of dust-aggregate growth in the planetesimal formation stage.

\subsection{Drag-force comparison with granular impact} \label{sec:drag-comparison}
The projectile motion in this study can be fitted well by Eq.~(\ref{eq:solution}), which has been proposed to explain the granular impact drag force. The two normalized fit parameters roughly become constant in the high-speed regime of our impacts~(Fig.~\ref{fig:k-d1}), but are velocity dependent in the low-speed regime. This variation probably comes from the elastic effect of the dust aggregates. As shown in Fig.~\ref{fig:Vep}(b), most of the impacts result in the rebound of the projectile, which is the elastic response of the compressed dust aggregate. The restitution coefficient depends on the impact energy (see Fig.~\ref{fig:Vep}(b)). When the impact energy is small, the restitution coefficient is relatively large. Therefore, the impacts possess an elastic as well as a plastic component. However, $k/({\pi}D)$ can basically be understood by considering the plastic deformation of cratering, as discussed in Sect.~\ref{sec:drag-force}. For small impact energies, the effect of elastic deformation is enhanced and added to plasticity-based $k/({\pi}D)$ value. This is a possible reason for the variation of $k/({\pi}D)$ in the small $v_0$ regime. This partially elastic effect might also strengthen the inertial drag. The pseudo decrease of $d_1{\rho_t}/{\rho_p}D$ could stem from this effect. From the experimental data, we consider $\epsilon{\lesssim}0.15$ is sufficient to apply the simple drag-force model with constant parameter values.

In impact into granular matter, $F_k$ originates from the frictional effect. In this study, however, the principal effect to cause $F_k$ corresponds to plastic deformation. Both frictional and plastic effects are dissipative. However, the form $F_k=kz$ is similar to the elastic behavior in terms of the equation of motion. In this sense, the drag force model Eq.~(\ref{eq:granular-drag}) should be improved. In fact, the value of $k$ in the granular impact cannot be fully explained by the simple Coulomb friction and hydrostatic pressure~\citep{Katsuragi:2007,Katsuragi:2013}. Usually, the dissipative effect can be modeled by a rate-dependent term in the equation of motion. While we have a $v^2$-dependent term in Eq.~(\ref{eq:granular-drag}), the $v$-dependent term is absent. Although the absence of the $v$-dependent term relates to the rate-dependent granular viscosity~\citep{Katsuragi:2016}, a $v$-dependent term improves the model in some cases~\citep{Nakamura:2013}. At present, Eq.~(\ref{eq:granular-drag}) can explain the motion of projectile very well both in granular and dust-aggregate impacts, at least as an empirical model. Furthermore, the form of $F_k$ is consistent with the energy scaling in the strength-dominant regime, Eq.~(\ref{eq:vol-scale}). To build more precise drag-force models, rate-dependent rheological properties should be studied in detail.

\section{Conclusion}
A simple experiment to characterize the dynamic mechanical properties of dust aggregates was performed. We dropped a solid spherical projectile onto a dust aggregate and measured the motion of the impinging projectile by using a high-speed camera. The experiment was performed in a vacuum chamber under the influence of gravity. From the acquired high-speed images, temporal developments of position $z$, velocity $v$, acceleration $a$ of the projectile, as well as the dynamic pressure $p_{\rm dy}$ were computed using image analysis. By varying the impact velocity and the density of the projectiles, the impact energy $E$ was varied over two orders of magnitude. In the relatively low impact-energy regime, the impact results in a spherical-indent cratering without any ejection of dust. In this cratering regime, rebound of the projectile can also be observed. When the impact energy is large enough, fragmentation of the target dust aggregate is caused. To break the target, the dynamic pressure must exceed the threshold value of $p_{\rm dy}^*{\simeq}10$~kPa. This critical dynamic pressure can be explained by the stress to open a crack. From the relation between crater volume $V_c$ and impact energy $E$, an effective yield strength $Y_{\rm k}=E/V_c=120$~kPa can be obtained, which relates to the plastic deformation of the target aggregate and whose value is consistent with the drag-based strength $k/({\pi}D)=200$~kPa. Here, the value of $k$ is computed from the fitting of kinematic data to Eq.~(\ref{eq:solution}). Another fit parameter in Eq.~(\ref{eq:solution}), $d_1$, indicates that the inertial drag is less important for dust-aggregate impacts, except for the very initial stage of the impact. From the relation between the maximum dynamic pressure max($p_{\rm dyn}$) and impact energy $E$, it turned out that the volume compressed by the impact $V_s=E/{\rm max(}p_{\rm dyn}$)${\simeq}4.6{\times}10^{-8}$~m$^{3}$ is close to the projectile volume $V_p$. These results provide fundamental understandings about mechanical properties of dust aggregates in the dynamic state. A simple model estimating the fragmentation threshold and the compaction degree of the impacted dust aggregate is proposed on the basis of our experimental result.

\acknowledgments
HK thanks JSPS KAKENHI Grant Nos.~15KK0158 and 15H03707 for financial support. JB thanks the Deutsche Forschungsgemeinschaft (DFG, grant Bl 298/24-1) and the Deutsches Zentrum f\"ur Luft- und Raumfahrt (DLR, grant 50WM1536) for financial support.

\bibliography{di}

\begin{thebibliography}{}
\expandafter\ifx\csname natexlab\endcsname\relax\def\natexlab#1{#1}\fi
\providecommand{\url}[1]{\href{#1}{#1}}

\bibitem[{Altshuler {et~al.}(2014)Altshuler, Torres, Gonz{\'a}lez-Pita,
  S{\'a}nchez-Colina, P{\'e}rez-Penichet, Waitukaitis, \&
  Hidalgo}]{Altshuler:2014}
Altshuler, E., Torres, H., Gonz{\'a}lez-Pita, A., {et~al.} 2014, Geophys. Res.
  Lett., 41, 3032

\bibitem[{Ambroso {et~al.}(2005)Ambroso, Santore, Abate, \&
  Durian}]{Ambroso:2005}
Ambroso, M.~A., Santore, C.~R., Abate, A.~R., \& Durian, D.~J. 2005, Phys. Rev.
  E, 71, 051305

\bibitem[{{Beitz} {et~al.}(2016){Beitz}, {Blum}, {Parisi}, \&
  {Trigo-Rodriguez}}]{Beitz:2016}
{Beitz}, E., {Blum}, J., {Parisi}, M.~G., \& {Trigo-Rodriguez}, J. 2016, \apj,
  824, 12

\bibitem[{{Beitz} {et~al.}(2013){Beitz}, {G{\"u}ttler}, {Nakamura},
  {Tsuchiyama}, \& {Blum}}]{Beitz:2013}
{Beitz}, E., {G{\"u}ttler}, C., {Nakamura}, A.~M., {Tsuchiyama}, A., \& {Blum},
  J. 2013, \icarus, 225, 558

\bibitem[{Bertini {et~al.}(2009)Bertini, Gutierrez, \& Sabolo}]{Bertini:2009}
Bertini, I., Gutierrez, P.~J., \& Sabolo, W. 2009, A{\&}A, 504, 625

\bibitem[{Blum \& Schr{\"a}pler(2004)}]{Blum:2004}
Blum, J., \& Schr{\"a}pler, R. 2004, Phys. Rev. Lett., 93, 115503

\bibitem[{Blum {et~al.}(2006)Blum, Schr{\"a}pler, Davidsson, \&
  Trigo-Rodriguez}]{Blum:2006}
Blum, J., Schr{\"a}pler, R., Davidsson, B. J.~R., \& Trigo-Rodriguez, J.~M.
  2006, ApJ, 652, 1768

\bibitem[{Blum \& Wurm(2008)}]{Blum:2008}
Blum, J., \& Wurm, G. 2008, Annu. Rev. Astro. Astrophys., 46, 21

\bibitem[{Blum {et~al.}(2014)Blum, Beitz, Bukhari, Gundlach, Hagemann,
  Hei{\ss}elmann, Kothe, Schr{\"a}pler, von Borstel, \& Weidling}]{Blum:2014}
Blum, J., Beitz, E., Bukhari, M., {et~al.} 2014, JoVE (Journal of Visualized
  Experiments), e51541

\bibitem[{Bukhari~Syed {et~al.}(2017)Bukhari~Syed, Blum, Jansson, \&
  Johansen}]{Syed:2017}
Bukhari~Syed, M., Blum, J., Jansson, K.~W., \& Johansen, A. 2017, ApJ, 834, 1

\bibitem[{de~Vet \& de~Bruyn(2007)}]{deVet:2007}
de~Vet, S.~J., \& de~Bruyn, J.~R. 2007, Phys. Rev. E, 76, 041306

\bibitem[{{Fulle} \& {Blum}(2017)}]{Fulle:2017}
{Fulle}, M., \& {Blum}, J. 2017, \mnras, 469, S39

\bibitem[{Guillard {et~al.}(2015)Guillard, Golshan, Shen, Vald{\`e}s, \&
  Einav}]{Guillard:2015}
Guillard, F., Golshan, P., Shen, L., Vald{\`e}s, J.~R., \& Einav, I. 2015, Nat.
  Phys., 11, 835

\bibitem[{G{\"u}ttler {et~al.}(2010)G{\"u}ttler, Blum, Zsom, Ormel, \&
  Dullemond}]{Guttler:2010}
G{\"u}ttler, C., Blum, J., Zsom, A., Ormel, C.~W., \& Dullemond, C.~P. 2010,
  A{\&}A, 513, A56

\bibitem[{G{\"u}ttler {et~al.}(2009)G{\"u}ttler, Krause, Geretshauser, Speith,
  \& Blum}]{Guttler:2009}
G{\"u}ttler, C., Krause, M., Geretshauser, R.~J., Speith, R., \& Blum, J. 2009,
  ApJ, 701, 130

\bibitem[{Heim {et~al.}(1999)Heim, Blum, Preuss, \& Butt}]{Heim:1999}
Heim, L.-O., Blum, J., Preuss, M., \& Butt, H.-J. 1999, Phys. Rev. Lett., 83,
  3328

\bibitem[{Herminghaus(2013)}]{Herminghaus:2013}
Herminghaus, S. 2013, {Wet granular matter}, A Truly Complex Fluid (World
  Scientific)

\bibitem[{Johnson(1985)}]{Johnson:1985}
Johnson, K.~L. 1985, {Contact Mechanics} (Cambridge: Cambridge University
  Press)

\bibitem[{Katsuragi(2010)}]{Katsuragi:2010}
Katsuragi, H. 2010, Phys. Rev. Lett., 104, 218001

\bibitem[{Katsuragi(2016)}]{Katsuragi:2016}
---. 2016, {Physics of Soft Impact and Cratering}, Vol. LNP 910 (Springer)

\bibitem[{Katsuragi \& Durian(2007)}]{Katsuragi:2007}
Katsuragi, H., \& Durian, D.~J. 2007, Nat. Phys., 3, 420

\bibitem[{Katsuragi \& Durian(2013)}]{Katsuragi:2013}
---. 2013, Phys. Rev. E, 87, 052208

\bibitem[{Meisner {et~al.}(2012)Meisner, Wurm, \& Teiser}]{Meisner:2012}
Meisner, T., Wurm, G., \& Teiser, J. 2012, A{\&}A, 544, A138

\bibitem[{Nakamura {et~al.}(2013)Nakamura, Setoh, Wada, Yamashita, \&
  Sangen}]{Nakamura:2013}
Nakamura, A.~M., Setoh, M., Wada, K., Yamashita, Y., \& Sangen, K. 2013,
  Icarus, 223, 222

\bibitem[{Nelson {et~al.}(2008)Nelson, Katsuragi, Mayor, \&
  Durian}]{Nelson:2008}
Nelson, E.~L., Katsuragi, H., Mayor, P., \& Durian, D.~J. 2008, Phys. Rev.
  Lett., 101, 068001

\bibitem[{Pacheco-V{\'a}zquez \& Ruiz-Su{\'a}rez(2011)}]{PachecoVazquez:2011}
Pacheco-V{\'a}zquez, F., \& Ruiz-Su{\'a}rez, J.~C. 2011, Phys. Rev. Lett., 107,
  218001

\bibitem[{Planes {et~al.}(2017)Planes, Mill{\'a}n, Urbassek, \&
  Bringa}]{Planes:2017}
Planes, M.~B., Mill{\'a}n, E.~N., Urbassek, H.~M., \& Bringa, E.~M. 2017,
  A{\&}A, doi:https://doi.org/10.1051/0004-6361/201730954, in press

\bibitem[{Poppe \& Schr{\"a}pler(2005)}]{Poppe:2005}
Poppe, T., \& Schr{\"a}pler, R. 2005, A{\&}A, 438, 1

\bibitem[{Royer {et~al.}(2011)Royer, Conyers, Corwin, Eng, \&
  Jaeger}]{Royer:2011}
Royer, J.~R., Conyers, B., Corwin, E.~I., Eng, P.~J., \& Jaeger, H.~M. 2011,
  EPL, 93, 28008

\bibitem[{Scheel {et~al.}(2008)Scheel, Seemann, Brinkmann, Di~Michiel,
  Sheppard, \& Herminghaus}]{Scheel:2008}
Scheel, M., Seemann, R., Brinkmann, M., {et~al.} 2008, J. Phys.: Condens.
  Matter, 20, 494236

\bibitem[{{Schr\"apler} {et~al.}(2017){Schr\"apler}, {Blum}, {Krijt}, \&
  {Raabe}}]{Schraepler:2017}
{Schr\"apler}, R., {Blum}, J., {Krijt}, S., \& {Raabe}, J.-H. 2017, submitted
  to ApJ

\bibitem[{Sirono(2004)}]{Sirono:2004}
Sirono, S.-i. 2004, Icarus, 167, 431

\bibitem[{Takehara {et~al.}(2010)Takehara, Fujimoto, \&
  Okumura}]{Takehara:2010}
Takehara, Y., Fujimoto, S., \& Okumura, K. 2010, EPL, 92, 44003

\bibitem[{Trigo-Rodriguez \& Llorca(2006)}]{TrigoRodriguez:2006}
Trigo-Rodriguez, J.~M., \& Llorca, J. 2006, Monthly Notices of the Royal
  Astronomical Society, 372, 655

\bibitem[{Uehara {et~al.}(2003)Uehara, Ambroso, Ojha, \& Durian}]{Uehara:2003}
Uehara, J.~S., Ambroso, M.~A., Ojha, R.~P., \& Durian, D.~J. 2003, Phys. Rev.
  Lett., 90, 194301

\bibitem[{Umbanhowar \& Goldman(2010)}]{Umbanhowar:2010}
Umbanhowar, P., \& Goldman, D.~I. 2010, Phys. Rev. E, 82, 010301

\bibitem[{Vald{\`e}s {et~al.}(2011)Vald{\`e}s, Fernandes, \&
  Einav}]{Valdes:2011}
Vald{\`e}s, J.~R., Fernandes, F.~L., \& Einav, I. 2011, Granular Matter, 14, 71

\bibitem[{Walsh {et~al.}(2003)Walsh, Holloway, Habdas, \&
  de~Bruyn}]{Walsh:2003}
Walsh, A.~M., Holloway, K.~E., Habdas, P., \& de~Bruyn, J.~R. 2003, Phys. Rev.
  Lett., 91, 104301

\bibitem[{{Weidling} {et~al.}(2009){Weidling}, {G{\"u}ttler}, {Blum}, \&
  {Brauer}}]{Weidling:2009}
{Weidling}, R., {G{\"u}ttler}, C., {Blum}, J., \& {Brauer}, F. 2009, \apj, 696,
  2036

\bibitem[{Wurm {et~al.}(2005)Wurm, Paraskov, \& Krauss}]{Wurm:2005}
Wurm, G., Paraskov, G., \& Krauss, O. 2005, Phys. Rev. E, 71, 021304

\bibitem[{Zsom {et~al.}(2010)Zsom, Ormel, G{\"u}ttler, Blum, \&
  Dullemond}]{Zsom:2010}
Zsom, A., Ormel, C.~W., G{\"u}ttler, C., Blum, J., \& Dullemond, C.~P. 2010,
  A{\&}A, 513, A57

\end{thebibliography}

\end{document}